\documentclass[aps,prb,twocolumn,floats]{revtex4}

\usepackage{bm}
\usepackage{graphicx}

\begin{document}

\title{CdSe/ZnSe quantum dot with a single Mn$^{2+}$ ion - a new system for a single spin manipulation}
\author{T. Smole\'nski}
\email{Tomasz.Smolenski@fuw.edu.pl}
\affiliation{
Institute of Experimental Physics, Faculty of Physics, University of Warsaw, ul. Ho\.za
69, 00-681 Warsaw, Poland.}

\date{\today}

\begin{abstract}
We present a magneto-optical study of individual self-assembled CdSe/ZnSe quantum dots doped with single Mn$^{2+}$ ions. Properties of the studied dots are analyzed analogously to more explored system of Mn-doped CdTe/ZnTe dots. Characteristic sixfold splitting of the neutral exciton emission line as well as its evolution in the magnetic field are described using a spin Hamiltonian model. Dynamics of both exciton recombination and Mn$^{2+}$ spin relaxation are extracted from a series of time-resolved experiments. Presence of a single dopant is shown not to affect the average excitonic lifetime measured for a number of nonmagnetic and Mn-doped dots. On the other hand, non-resonant pumping is demonstrated to depolarize the Mn$^{2+}$ spin in a quantum dot placed in external magnetic field. This effect is utilized to determine the ion spin relaxation time in the dark.
\end{abstract}

\maketitle

\section{Introduction}

Semiconductor quantum dots (QDs) with single magnetic ions have recently been a subject of intense research activity as systems close to the ultimate limit of miniaturization of magnetic memory\cite{Besombes_PRL_2004, Leger_PRL_2005, Leger_PRL_2006, Kudelski_PRL_2007, LeGall_PRL_2009, Goryca_PRL_2009, Reiter_PRL_2009, LeGall_PRB_2010, krebs_PRL_2011}. They provide an exceptional opportunity to integrate the QD optical properties with individual spin of magnetic dopant. One of its most important consequences is the possibility of optical control over the ion spin state resulting from the exchange interaction between a neutral exciton confined in a dot and the ion spin. The efficient spin readout at the moment of exciton recombination is provided by the polarization and energy of emitted photon\cite{Besombes_PRL_2004}. On the other hand, an optical manipulation of the ion spin is possible through injection of spin-polarized excitons to the dot\cite{LeGall_PRL_2009, Goryca_PRL_2009, LeGall_PRB_2010}. Importantly, the single magnetic impurity embedded in the QD may be also exploited as a spin memory exhibiting relatively long spin relaxation time\cite{LeGall_PRL_2009, Goryca_PRL_2009}.

Despite the outstanding properties and possible applications, a serious limitation for operation of QDs with single magnetic dopants was typically attributed to the non-radiative recombination channel introduced by magnetic ion when the exciton energy exceeds the intra-ionic transition energy. In particular, the presence of the ion was commonly believed to quench the exciton emission for a wide group of systems including self-assembled CdSe/ZnSe QDs. Such a belief was predominantly based on a direct observation of efficient photoluminescence (PL) quenching in case of bulk diluted magnetic semiconductors or QDs doped with a large number of magnetic ions\cite{Oka_JLum_1999, Lee_PRB_2005, Chekhovich_PRB_2007, Beaulac_NanoLett_2008, Beaulac_AdvFunMat_2008, Bussian_NatMat_2009, Pacuski_PRB_2011, Papaj_JCG_2014}. As a result, the only dots with single magnetic dopants explored so far were CdTe/ZnTe\cite{Besombes_PRL_2004, Leger_PRL_2005, Leger_PRL_2006, LeGall_PRL_2009, Goryca_PRL_2009, LeGall_PRB_2010} and InAs/GaAs\cite{Kudelski_PRL_2007, Krebs_PRB_2009, krebs_PRL_2011} QDs containing Mn$^{2+}$ ions, for which the exciton energy is lower than the intra-ionic transition energy.

Very recently a fabrication of two qualitatively new QD systems has been reported, namely CdTe/ZnTe dots with single Co$^{2+}$ ions and CdSe/ZnSe QDs with single Mn$^{2+}$ ions\cite{Kobak_2014}. Despite being energetically allowed, the ion-related emission quenching has been demonstrated to be negligible in both systems by observation of similar excitonic lifetimes for a reference undoped QD and a dot containing a single magnetic ion. This finding extends the possibility of optical studies of QDs with individual magnetic impurities for a wide group of semiconductor materials and transition metal ions. Crucially, it enables tuning the properties of such QDs in a desired way. As an example, it has been shown that switching from CdTe to CdSe dot material results in significant prolongation of a single Mn$^{2+}$ ion spin relaxation time\cite{Kobak_2014}.

Here we summarize the results of the detailed optical study of CdSe/ZnSe quantum dots doped with single Mn$^{2+}$ ions. In particular, we perform a comprehensive analysis of the evolution of neutral exciton emission spectrum in the magnetic field applied in Faraday geometry within the frame of a spin Hamiltonian model. Moreover, we provide more robust evidence for inefficiency of the ion-related emission quenching by comparing the excitonic lifetimes measured for several randomly selected dots, both nonmagnetic and containing single ions. Finally, we present experimental results indicating that Mn$^{2+}$ ion spin in a CdSe QD can be optically influenced by non-resonant excitation in the presence of external magnetic field. This effect underlies the idea of our time-resolved measurements of the ion spin relaxation dynamics, which results are thoroughly discussed.

\section{Sample and experimental setup}

The studied sample contains a single layer of molecular beam epitaxy grown self-organized CdSe/ZnSe QDs. The dots are weakly doped with manganese ions. Their concentration is optimized to maximize the probability of formation of QDs with exactly one Mn$^{2+}$ ion in each dot. Detailed description of the sample growth procedure can be found in Ref. \onlinecite{Kobak_2014}.

Our measurements are performed using a micro-photoluminescence setup. The sample is placed inside a magneto-optical helium-bath cryostat and cooled down to temperature of about 1.6 K. The cryostat is equipped with a superconducting magnet producing magnetic field up to 10 T. The field is applied in Faraday configuration, parallel to the growth axis of the sample. The PL is excited non-resonantly at 405 nm using either a continues-wave diode laser, or a frequency doubled titanium-sapphire laser emitting 2 ps pulses with repetition frequency of 76 MHz. High spatial resolution is achieved in our experiments by using a reflective-type microscope objective attached directly to the sample surface. It enables to focus laser beam to spot size of diameter not exceeding 1 $\mu$m. The luminescence of the QDs is spectrally resolved using 0.5 m monochromator. The set of polarization optics (including a linear polarizer, $\lambda/2$, and $\lambda/4$ waveplates) is placed in the signal beam in order to detect PL with polarization resolution. 

The time-integrated PL spectra are recorded with a CCD camera. In case of time-resolved experiments, two different techniques are employed. The measurements of radiative lifetimes are carried out under pulsed excitation using a low-jitter avalanche photodiode (APD) or a synchroscan Hamamatsu streak camera. The overall temporal resolution in both cases is up to 40 ps and 10 ps, respectively. The second experimental technique is used for time-resolved measurements of Mn$^{2+}$ ion spin relaxation dynamics. In this case, an acousto-optic modulator is utilized to repeatedly switch on and off the cw excitation (with switching time below 10 ns). The modulator is driven by a generator enabling to set both the switching frequency and the length of the dark period. Temporal profiles of the PL signal under such excitation are recorded with a high efficiency APD providing sub-nanosecond temporal resolution.

\section{Magneto-PL of CdSe QD with a single Mn$^{2+}$ ion}
\label{sec:magneto}

\begin{figure}
\includegraphics{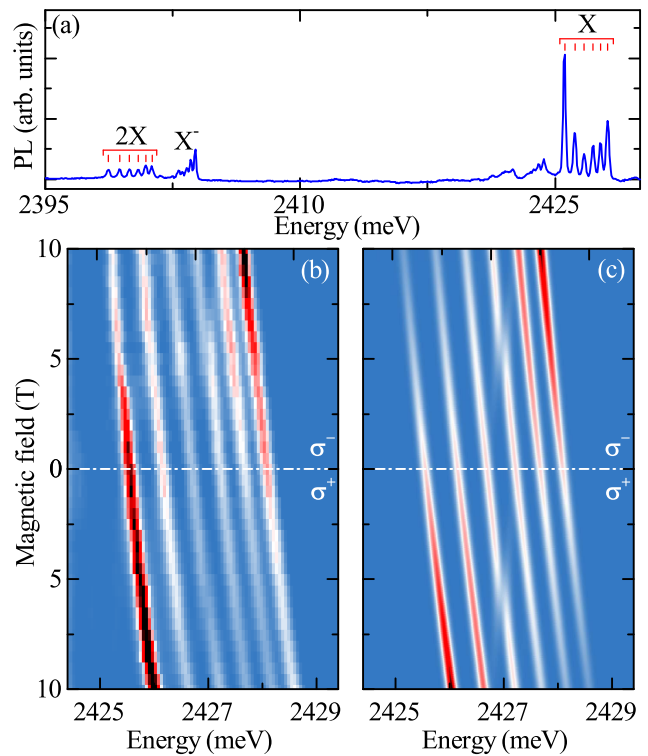}
\caption{(Color online) (a) Zero-field PL spectrum of a single CdSe/ZnSe QD containing individual Mn$^{2+}$ ion ($T=1.6$ K). (b), (c) Color-scale plots presenting the measured (b) and simulated (c) evolution of neutral exciton PL spectrum in the magnetic field in Faraday geometry. The upper (lower) parts of the plots correspond to spectra detected in $\sigma^-$ ($\sigma^+$) polarization.
 \label{fig1}}
\end{figure}

Micro-photoluminescence spectra of small QD ensembles (down to a hundred of dots within the laser spot) exhibit an inhomogeneous broadening with a characteristic line structure. Relatively low density of QDs in our sample enables us to distinguish well isolated groups of emission lines originating from single QDs in the low energy tail of the PL band. It is possible to find a number of dots, both nonmagnetic and containing individual Mn$^{2+}$ ions. An example PL spectrum of Mn-doped QD is shown in Fig. \ref{fig1}(a). The emission lines are identified as originating from recombination of neutral exciton (X), negatively charged trion (X$^-$), and biexciton (2X), as marked in Fig. \ref{fig1}(a). Each of them is split by the exchange interaction between carriers and magnetic ion. Focusing on the neutral exciton, its emission exhibits characteristic sixfold splitting arising mainly due to the heavy hole-Mn exchange. Since the total spin of the Mn$^{2+}$ ion is equal to 5/2, it has six possible projections on the growth axis, which defines the quantization axis of the hole. As a result, we observe six emission lines of the neutral exciton, each of them corresponding to a specific ion spin projection for a given circular polarization of detection. 

In order to additionally confirm the presence of the ion in the studied dot and thoroughly analyze its interaction with the neutral exciton, we measure the evolution of X PL spectrum in magnetic field with polarization resolution. It is shown in Fig. \ref{fig1}(b). The excitonic Zeeman effect introduced by the magnetic field splits each of X emission lines. Consequently, we observe twelve lines at non-zero field, six in each circular polarization. On the other hand, the emission energies are not affected by the Zeeman splitting of Mn$^{2+}$ ion states, since it is identical in the initial and final states of X optical transitions. Nevertheless, this splitting influences the line intensities. More specifically, it leads to thermalisation of the Mn$^{2+}$ ion spin towards $-5/2$ state, which manifest itself as an increased intensity of high-energy (low-energy) line in $\sigma^-$ ($\sigma^+$) circular polarization, clearly visible in Fig. \ref{fig1}(b) at high magnetic field. However, this effect does not explain the asymmetry appearing in the distribution of X emission intensities at $B=0$. We interpret it as arising due to X-Mn spin relaxation during the exciton lifetime, which leads to an increased intensities of lower-energy lines. The presence of such relaxation affects also X PL spectra at $B>0$. In such case, the intensities are governed by an interplay of X-Mn spin relaxation and Mn$^{2+}$ spin thermalisation in the empty dot. As a result, the field-induced enhancement of low-energy line intensity in $\sigma^+$ polarization is more pronounced compared to the increase of high-energy line intensity in $\sigma^-$ polarization (Fig. \ref{fig1}(b)).

A quantitative description of X magneto-PL is performed within a frame of a simple model of the neutral exciton inside Mn-doped QD previously exploited for CdTe/ZnTe QDs\cite{Besombes_PRL_2004, Goryca_PRB_2010}. It is given by the following Hamiltonian:
\begin{eqnarray}
\mathcal{H} &=& g_{Mn}\mu_BBS_z+g_{h}\mu_BBJ_z+g_{e}\mu_BB\sigma_z\nonumber\\
&+&I_{h}\vec{S}\vec{J}+I_{e}\vec{S}\vec{\sigma}-\frac{2\delta_0}{3}J_z\sigma_z+\frac{2\delta_1}{3}\left(J_x^3\sigma_x-J_y^3\sigma_y\right)\nonumber\\
&+&\Delta \left(J_x^2+J_y^2\right)+\rho\left(J_x^2-J_y^2\right)-\eta S_z^2+\gamma B^2,
\label{eq:ham}
\end{eqnarray}
where $B$ denotes the magnetic field parallel to the growth axis $z$, while $\vec{S}$, $\vec{J}$ and $\vec{\sigma}$ are Mn$^{2+}$ ion, hole and electron spin operators, respectively. The first three terms represent the Zeeman energies of the ion, the hole and the electron with corresponding g-factors $g_{Mn}$, $g_h$ and $g_e$ (the Mn$^{2+}$ g-factor is assumed to be 2.0 basing on the literature data\cite{Title_PR_1963}). $I_h$ and $I_e$ are the hole-Mn and electron-Mn exchange integrals. The isotropic and anisotropic parts of electron-hole exchange are described by the effective energies $\delta_0$ and $\delta_1$\cite{Hommel_PRL_1999, Bayer_PRB_2002,Akimov_PRB_2005}. Furthermore, $\Delta$ is the heavy-light hole energy splitting and $\rho$ represents the strength of the valence band mixing (VBM). The second-to-last term $-\eta S_z^2$ describes the perturbation of X-Mn energy levels imposed by the configuration mixing\cite{Trojnar_PRL_2011, Trojnar_PRB_2013}. Finally, $\gamma$ is the excitonic diamagnetic shift constant. We note that the first term in $\mathcal{H}$ is also the Hamiltonian of the single ion in the empty dot (after X recombination).

Numerical diagonalization of the above introduced Hamiltonian enables direct calculation of the excitonic optical transition energies in two circular polarizations. By adjusting the model parameters we are able to fully reproduce the evolution of X emission energies in the magnetic field, as shown in Fig. \ref{fig1}(c). The presented PL intensities are computed as a product of oscillator strength and Boltzmann distribution of the ion spin state occupancies for an effective temperature $T=30$ K corresponding to Mn$^{2+}$ spin thermalisation in the empty dot. Since our model does not take into account the existence of X-Mn spin relaxation during the exciton lifetime, the calculated intensities are not fully consistent with the experimental results.

In order to provide a detailed information about the obtained Hamiltonian parameters, we now discuss the most characteristic features of X PL evolution in the magnetic field. Focusing first on the zero-field spectrum, we observe a small, but not negligible irregularity in energy spacing of X emission lines. In particular, the splitting of two consecutive lines is clearly decreasing with the line energy. Following previous studies of CdTe Mn-doped QDs\cite{Trojnar_PRL_2011, Trojnar_PRB_2013}, we interpret this effect as arising due to the configuration mixing between ground and excited excitonic states. It is directly described by $-\eta S_z^2$ term in the Hamiltonian, which enables us to determine $\eta\simeq23\ \mu$eV, which is similar to the values reported for CdTe dots\cite{Besombes_PRB_2014}.

At magnetic field of about 6 T, a pronounced anticrossing of the middle-energy emission lines is clearly visible in both circular polarizations. Each of these lines corresponds to the same $-1/2$ ion spin projection, but different spin states of bright excitons. As a result, the excitonic Zeeman effect is progressively reducing the zero-field splitting of the two lines related to the exchange interaction with the Mn$^{2+}$ ion. Finally, at 6 T their energies are brought into coincidence and anisotropic part of the electron-hole exchange interaction starts to mix the excitonic states with opposite spins,  which leads to the observed anticrossing (analogous effect was also demonstrated for CdTe QDs\cite{Leger_PRL_2005, Goryca_PRB_2010}). The splitting of emission lines at the anticrossing is directly equal to $\delta_1$, which is determined to be about 150 $\mu$eV for the studied dot. Similar anticrossings are expected to occur for the emission lines corresponding to $-3/2$ and $-5/2$ ion spin projections at magnetic fields of 17 T and 28 T, which are however inaccessible experimentally in the present study. 

Coherently with previously published results in Ref. \onlinecite{Kobak_2014}, we do not observe any signature of the dark exciton optical transitions occurring with simultaneous flip of the Mn$^{2+}$ spin. They should be partially allowed by off-diagonal terms of electron-Mn exchange interaction or hole-Mn exchange accompanied by VBM, as it was previously shown for CdTe QDs with single Mn$^{2+}$ ions\cite{Besombes_PRL_2004, Goryca_PRB_2010}. The existence of both mechanisms in the case of CdSe Mn-doped QDs is strongly indicated by non-zero $s,p-d$ exchange constants $N_0\alpha$ and $N_0\beta$ in bulk (Cd,Mn)Se\cite{Arciszewska_JChemSol_1966} and significant VBM typically observed for nonmagnetic CdSe dots\cite{Koudinov_PRB_2004}. However, their efficiency might be much lower compared to CdTe QDs due to increased energy splitting $\delta_0$ of bright and dark excitons related to the isotropic part of electron-hole exchange interaction. More specifically, the previous studies of CdSe QDs revealed $\delta_0$ to be almost 2 meV\cite{Akimov_PRB_2005}, which is more than two times larger than $\delta_0$ determined for CdTe QDs\cite{tkaz_prb_2011, Smolenski_PRB_2012}. As a result, the observation of anticrossings of bright and dark exciton transitions for CdSe Mn-doped dots would presumably require application of stronger magnetic field needed to compensate larger $\delta_0$ energy splitting. The lack of the dark exciton fingerprint in the present experimental data prevents us to independently determine the electron-Mn and hole-Mn exchange integrals as well as both carriers g-factors. Moreover, we are not able to precisely obtain VBM strength corresponding to the ratio of $\rho$ and $\Delta$ parameters in the Hamiltonian. We note only that the bright exciton g-factor $3g_h-g_e$ corresponds to about 1.6, while the zero-field splitting of X sextuplet $\frac{5}{2}(3I_h-I_e)$ is equal to 2.54 meV (the formulas are given in the absence of VBM).

\section{X lifetime in nonmagnetic and Mn-doped CdSe QD}

\begin{figure}
\includegraphics{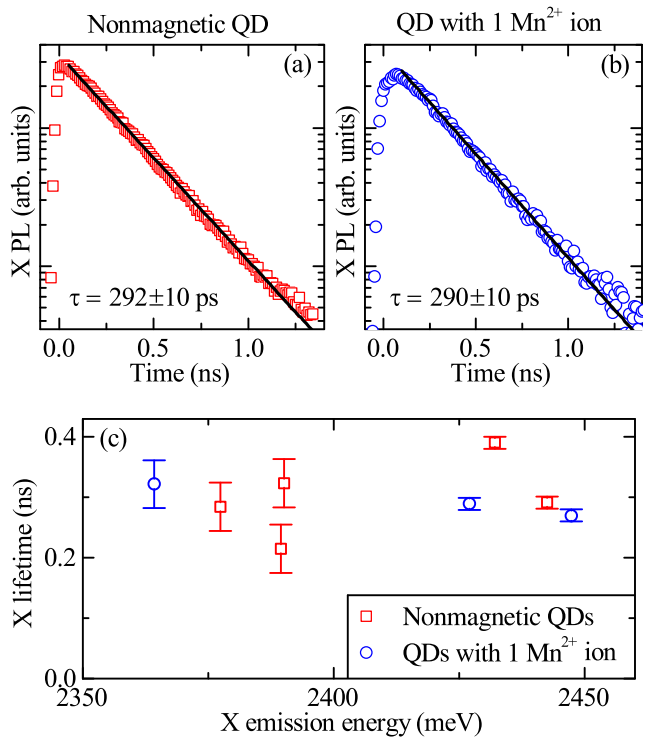}
\caption{(Color online) (a),(b) Neutral exciton PL decay profiles measured for a nonmagnetic QD (a) and a dot containing a single Mn$^{2+}$ ion  (b) at zero magnetic field and temperature of 1.6 K. To account for the background emission, the data are corrected by subtraction of a reference signal measured at emission energy corresponding to flat PL background. Solid lines represent fits of single exponential decays with indicated X lifetime values. (c) Correlation of X lifetime and emission energy for Mn-doped and undoped QDs. Each experimental point corresponds to a different randomly selected dot.
 \label{fig2}}
\end{figure}

The investigated CdSe/ZnSe QDs are characterized by emission energies ranging from about 2.3 eV to 2.5 eV. Since the manganese intra-ionic transition energy corresponds to 2.2 eV, the exciton non-radiative recombination leading to excitation of the Mn$^{2+}$ ion is energetically allowed for Mn-doped dots. However, the inefficiency of this recombination channel was revealed by similar exciton decay times previously determined for a QD with and without a single ion in Ref. \onlinecite{Kobak_2014}. In order to provide an additional confirmation of this finding, we compare X lifetimes measured for a number of dots, both nonmagnetic and Mn-doped. Our time-resolved experiments are carried out under sufficiently low excitation power assuring negligible influence of the biexciton cascaded emission on X PL decay profile. In the case of dots with single Mn$^{2+}$ ions, the presence of aforementioned X-Mn spin relaxation strongly affects the PL temporal profiles corresponding to different exchange-split X emission lines. Since their detailed analysis remains beyond the scope of this paper, we focus on a decay profile determined for the low-energy line.

The X PL temporal profiles obtained for an exemplary nonmagnetic and Mn-doped dot are presented in Figs. \ref{fig2}(a) and \ref{fig2}(b), respectively. They exhibit monoexponential character, which enables us to directly determine the corresponding X lifetimes. Their values measured for several, randomly selected dots are shown in Fig. \ref{fig2}(c) versus X emission energy. We observe no correlation between these two quantities both for nonmagnetic and Mn-doped QDs. Most importantly, the average X lifetime of $300\pm60$ ps in the former case is almost equal to the mean lifetime of $295\pm30$ ps determined for dots containing individual ions. This result clearly shows that introduction of a single Mn$^{2+}$ ion to a CdSe/ZnSe QD does not reduce the excitonic decay time, confirming the ineffectiveness of the ion-related X emission quenching. Following Ref. \onlinecite{Kobak_2014}, we interpret this as arising due to a discrete density of states of a QD with an individual magnetic impurity. In such case, even if non-radiative X recombination related to the intra-ionic transition is energetically allowed, it is not efficient due to a possible energy mismatch.

\section{Mn$^{2+}$ spin relaxation dynamics in CdSe QD}

\begin{figure}
\includegraphics{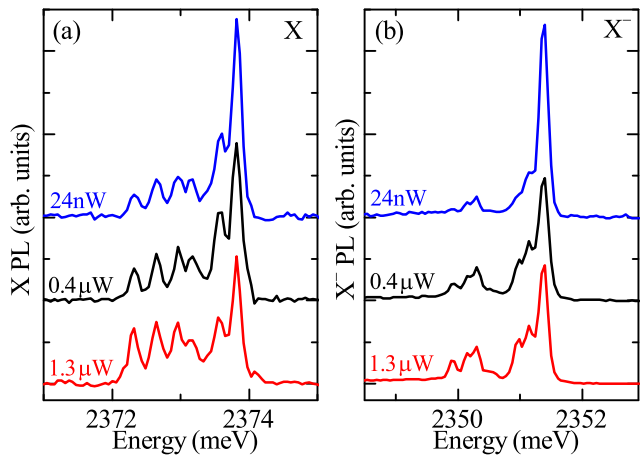}
\caption{(Color online) The neutral exciton (a) and negatively charged exciton (b) PL spectra measured for indicated excitation powers at magnetic field of 4 T and temperature of 1.6 K. The spectra are detected in $\sigma^-$ polarization and normalized (i.e., divided by the total integrated intensity).
 \label{fig3}}
\end{figure}

In order to measure Mn$^{2+}$ spin relaxation time in a CdSe QD, we first need to establish an optical method enabling manipulation of the ion spin state. According to the previous studies of CdTe QDs\cite{LeGall_PRL_2009, Goryca_PRL_2009}, it might be achieved through injection of spin-polarized excitons to the dot, which act on the Mn$^{2+}$ ion via the exchange interaction and orient its spin. However, application of this technique would require identification of a spin-conserving optical excitation channel for the studied dots. Instead, we utilize a much simpler approach also demonstrated for CdTe dots\cite{Besombes_PRL_2004}. Its underlying idea is based on depolarization of the Mn$^{2+}$ spin via exchange interaction with spin-unpolarized excitons, which can be easily injected to the dot under non-resonant excitation. In order for this mechanism to become effective, the ion spin should be initially polarized by application of the magnetic field in Faraday configuration. In such case, as it was previously mentioned in section \ref{sec:magneto}, the Mn$^{2+}$ Zeeman splitting leads to thermalisation of the ion spin towards the state of alignment to the field direction (i.e., $-5/2$ state for $B>0$). When the excitons of random spins are injected to the dot, the effective temperature of the Mn$^{2+}$ ion is increased leading to the partial depolarization of the ion spin. The presence of this effect is clearly visible in Fig. \ref{fig3}(a) presenting X PL spectra at $B=4$ T measured under different powers of non-resonant excitation in $\sigma^-$ polarization of detection. At low-power regime, high-energy line corresponding to $-5/2$ ion spin projection is dominating the spectrum. However, nonuniform distribution of six emission lines intensities progressively disappears as the excitation power is increased. It reflects the change of the ion spin state induced by injection of excitons to the dot. Since the Mn$^{2+}$ spin can be independently probed by different excitonic complexes, the same effect should be also observed in the case of negatively charged exciton. This prediction is fully reproduced by the power-dependence of X$^{-}$ PL spectra measured at the same magnetic field, which is shown in Fig. \ref{fig3}(b). It unequivocally confirms that the observed change of emission lines intensities is solely related to depolarization of the ion spin.

\begin{figure}
\includegraphics{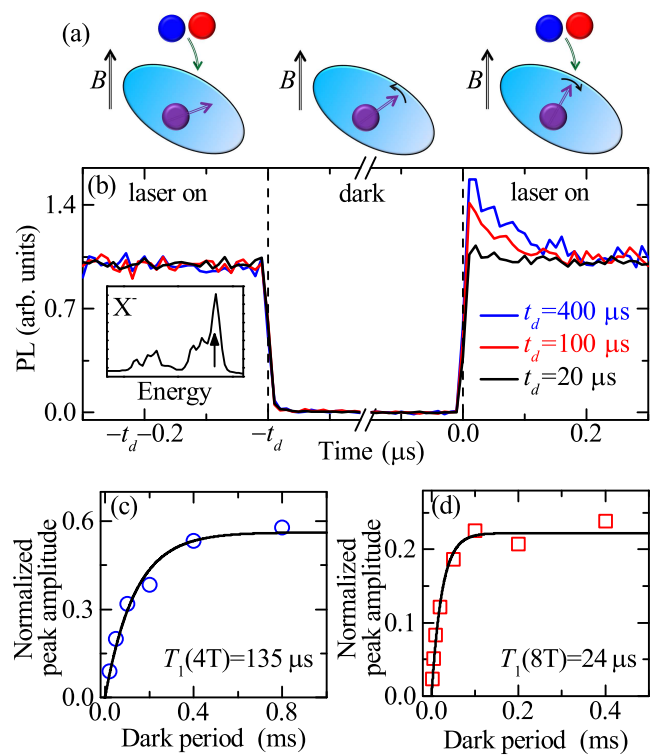}
\caption{(Color online) (a) Scheme describing the excitation sequence used in the time-resolved experiment and corresponding evolution of the ion spin. (b) Time-dependent PL intensity of high-energy X$^-$ emission line (indicated at inset) detected in $\sigma^-$ polarization at $B=4$ T for different lengths of dark period $t_d$ (measurements are performed at $T=1.6$ K). (c), (d) The normalized amplitude of PL intensity peak obtained just after switching on the excitation as a function of the length of the dark period for $B=4$ T (c) and $B=8$ T (d). Solid lines represent the fits of exponential curves with indicated Mn$^{2+}$ spin relaxation times.
 \label{fig4}}
\end{figure}

Taking advantage of the possibility to optically influence Mn$^{2+}$ spin in a CdSe QD, we measure the ion spin relaxation dynamics in a time-resolved experiment, in which the optical excitation is on/off modulated. The idea of the experiment is schematically described in Fig. \ref{fig4}(a). The Mn-doped QD is placed in external magnetic field. First, we depolarize the ion spin using non-resonant excitation. Then the laser is switched off for a controlled period of time. During this time the QD remains in the dark and the Mn$^{2+}$ spin relaxation processes lead to a progressive alignment of the ion spin towards the direction of magnetic field. In order to perform the Mn$^{2+}$ spin readout, the laser is finally switched on again. Described excitation sequence is subsequently repeated with appropriately low frequency assuring that the Mn$^{2+}$ spin reaches a steady state each time when the laser is on. During the experiment, the ion spin is monitored by measuring the PL intensity of high-energy X$^-$ emission line in $\sigma^-$ polarization of detection.

Exemplary temporal profiles obtained for different lengths of the dark period at magnetic field of 4 T are shown in Fig. \ref{fig4}(b). The selected emission line corresponds to the ion state with spin aligned towards the field direction. Since the probability of finding the ion in this state is increased by spin relaxation occurring in the dark period, the pronounced PL intensity peak is observed just after switching on the laser. The peak vanishes within about 100 ns due to depolarization of the magnetic ion spin induced by injection of excitons to the QD. Importantly, the peak amplitude clearly increases for longer dark period. This is directly reflecting the degree of the ion spin relaxation during this dark time. Consequently, we can use a normalized peak amplitude as a quantitative measure of the spin relaxation. Its dependence on the length of the dark period measured at magnetic field of 4 T and 8 T is shown in Figs. \ref{fig4}(c) and \ref{fig4}(d). The experimental results are fitted with exponential curves. On this basis we determine the Mn$^{2+}$ ion spin relaxation time in the dark at $B=4$ T and $B=8$ T to be equal to 135 $\mu$s and 24 $\mu$s, respectively. Both of these times are over an order of magnitude longer compared to respective relaxation times of the Mn$^{2+}$ ion spin in a CdTe QD\cite{Kobak_2014}. This finding clearly shows that switching to the QD material with weaker spin-orbit interaction significantly increases the storage time of information on the ion spin. From this point of view CdSe/ZnSe QDs with single Mn$^{2+}$ ions are more promising in possible future applications as optically controlled spin memories than CdTe/ZnTe QDs.

It is noteworthy that determined Mn$^{2+}$ spin relaxation time for the CdSe QD is much longer for lower magnetic field. Similar trend was previously observed in the case of CdTe QDs, where the ion spin relaxation time is about 0.4 ms at $B=1$ T (Ref. \onlinecite{Goryca_PRL_2009}) and 5 $\mu$s at $B=4$ T (Ref. \onlinecite{Kobak_2014}). On this basis, one could expect even longer Mn$^{2+}$ spin relaxation time at $B=1$ T for the currently studied system. However, at such field the Zeeman splitting of the ion states is rather weak, which results in a small field-induced ion spin orientation. As a consequence, our technique of spin relaxation measurement basing on optically-induced depolarization of the Mn$^{2+}$ spin becomes ineffective. Instead, the ion spin relaxation rate at the low-field regime could be possibly determined employing the method utilizing optical orientation of the Mn$^{2+}$ spin mediated by injection of spin-polarized excitons\cite{LeGall_PRL_2009, Goryca_PRL_2009}. Further experimental studies are needed to demonstrate the applicability of this technique in the case of CdSe QDs.

\section{Summary}

We have presented extended optical study of CdSe/ZnSe QDs containing single Mn$^{2+}$ ions with respect to our recent work published in Ref. \onlinecite{Kobak_2014}. The identification of dots with exactly one ion was based on the observation of characteristic sixfold splitting of neutral exciton emission line as well as on the analysis of evolution of X emission energies in external magnetic field applied along the growth axis. We have demonstrated that measured evolution can be fully reproduced by a simple spin Hamiltonian model previously proposed for CdTe/ZnTe QDs with single Mn$^{2+}$ ions. The obtained model parameters describing X-Mn complex in a CdSe QD were thoroughly discussed with indication of similarities and differences between CdTe and CdSe dots. 

Our time-resolved measurements of the neutral exciton lifetime performed for a number of randomly selected CdSe QDs revealed similar average lifetime for nonmagnetic and Mn-doped dots. This finding enabled us to confirm that the optical properties of a CdSe QD with exactly one ion are not affected by excitonic emission quenching related to the intra-ionic transitions.

We have also investigated a prospect of utilizing Mn-doped CdSe QD as an optically controlled spin memory. In particular, we have directly demonstrated an all-optical method of the ion spin depolarization by non-resonant excitation of a QD placed in external magnetic field. Using this method, we measured the ion spin relaxation dynamics in a time-resolved experiment with modulated optical excitation. The ion spin relaxation times in the dark at magnetic field of 4 T and 8 T were found to be over an order of magnitude longer compared to respective relaxation times of the Mn$^{2+}$ ion spin in previously studied CdTe QD.

\begin{acknowledgments}
The author thanks Tomasz Kazimierczuk, Tomasz Jakubczyk, Piotr Kossacki, Micha\l{} Nawrocki and Wojciech Pacuski for help in manuscript preparation. This work was supported by the Polish Ministry of Science and Higher Education in years 2012-2016 as research grant ''Diamentowy Grant'', by the Polish National Science Centre under decisions DEC-2011/02/A/ST3/00131, DEC-2012/07/N/ST3/03130, DEC-2013/09/B/ST3/02603, by the Polish National Centre for Research and Development project LIDER, and by Foundation for Polish Science (FNP) subsidy ''Mistrz''.
\end{acknowledgments}


\begin{thebibliography}{31}
\expandafter\ifx\csname natexlab\endcsname\relax\def\natexlab#1{#1}\fi
\expandafter\ifx\csname bibnamefont\endcsname\relax
  \def\bibnamefont#1{#1}\fi
\expandafter\ifx\csname bibfnamefont\endcsname\relax
  \def\bibfnamefont#1{#1}\fi
\expandafter\ifx\csname citenamefont\endcsname\relax
  \def\citenamefont#1{#1}\fi
\expandafter\ifx\csname url\endcsname\relax
  \def\url#1{\texttt{#1}}\fi
\expandafter\ifx\csname urlprefix\endcsname\relax\def\urlprefix{URL }\fi
\providecommand{\bibinfo}[2]{#2}
\providecommand{\eprint}[2][]{\url{#2}}

\bibitem[{\citenamefont{Besombes et~al.}(2004)\citenamefont{Besombes, L\'eger,
  Maingault, Ferrand, Mariette, and Cibert}}]{Besombes_PRL_2004}
\bibinfo{author}{\bibfnamefont{L.}~\bibnamefont{Besombes}},
  \bibinfo{author}{\bibfnamefont{Y.}~\bibnamefont{L\'eger}},
  \bibinfo{author}{\bibfnamefont{L.}~\bibnamefont{Maingault}},
  \bibinfo{author}{\bibfnamefont{D.}~\bibnamefont{Ferrand}},
  \bibinfo{author}{\bibfnamefont{H.}~\bibnamefont{Mariette}}, \bibnamefont{and}
  \bibinfo{author}{\bibfnamefont{J.}~\bibnamefont{Cibert}},
  \bibinfo{journal}{Phys. Rev. Lett.} \textbf{\bibinfo{volume}{93}},
  \bibinfo{pages}{207403} (\bibinfo{year}{2004}).

\bibitem[{\citenamefont{L\'eger et~al.}(2005)\citenamefont{L\'eger, Besombes,
  Maingault, Ferrand, and Mariette}}]{Leger_PRL_2005}
\bibinfo{author}{\bibfnamefont{Y.}~\bibnamefont{L\'eger}},
  \bibinfo{author}{\bibfnamefont{L.}~\bibnamefont{Besombes}},
  \bibinfo{author}{\bibfnamefont{L.}~\bibnamefont{Maingault}},
  \bibinfo{author}{\bibfnamefont{D.}~\bibnamefont{Ferrand}}, \bibnamefont{and}
  \bibinfo{author}{\bibfnamefont{H.}~\bibnamefont{Mariette}},
  \bibinfo{journal}{Phys. Rev. Lett.} \textbf{\bibinfo{volume}{95}},
  \bibinfo{pages}{047403} (\bibinfo{year}{2005}).

\bibitem[{\citenamefont{L\'eger et~al.}(2006)\citenamefont{L\'eger, Besombes,
  Fern\'andez-Rossier, Maingault, and Mariette}}]{Leger_PRL_2006}
\bibinfo{author}{\bibfnamefont{Y.}~\bibnamefont{L\'eger}},
  \bibinfo{author}{\bibfnamefont{L.}~\bibnamefont{Besombes}},
  \bibinfo{author}{\bibfnamefont{J.}~\bibnamefont{Fern\'andez-Rossier}},
  \bibinfo{author}{\bibfnamefont{L.}~\bibnamefont{Maingault}},
  \bibnamefont{and} \bibinfo{author}{\bibfnamefont{H.}~\bibnamefont{Mariette}},
  \bibinfo{journal}{Phys. Rev. Lett.} \textbf{\bibinfo{volume}{97}},
  \bibinfo{pages}{107401} (\bibinfo{year}{2006}).

\bibitem[{\citenamefont{Kudelski et~al.}(2007)\citenamefont{Kudelski,
  Lema\^{i}tre, Miard, Voisin, Graham, Warburton, and
  Krebs}}]{Kudelski_PRL_2007}
\bibinfo{author}{\bibfnamefont{A.}~\bibnamefont{Kudelski}},
  \bibinfo{author}{\bibfnamefont{A.}~\bibnamefont{Lema\^{i}tre}},
  \bibinfo{author}{\bibfnamefont{A.}~\bibnamefont{Miard}},
  \bibinfo{author}{\bibfnamefont{P.}~\bibnamefont{Voisin}},
  \bibinfo{author}{\bibfnamefont{T.~C.~M.} \bibnamefont{Graham}},
  \bibinfo{author}{\bibfnamefont{R.~J.} \bibnamefont{Warburton}},
  \bibnamefont{and} \bibinfo{author}{\bibfnamefont{O.}~\bibnamefont{Krebs}},
  \bibinfo{journal}{Phys. Rev. Lett.} \textbf{\bibinfo{volume}{99}},
  \bibinfo{pages}{247209} (\bibinfo{year}{2007}).

\bibitem[{\citenamefont{Le~Gall et~al.}(2009)\citenamefont{Le~Gall, Besombes,
  Boukari, Kolodka, Cibert, and Mariette}}]{LeGall_PRL_2009}
\bibinfo{author}{\bibfnamefont{C.}~\bibnamefont{Le~Gall}},
  \bibinfo{author}{\bibfnamefont{L.}~\bibnamefont{Besombes}},
  \bibinfo{author}{\bibfnamefont{H.}~\bibnamefont{Boukari}},
  \bibinfo{author}{\bibfnamefont{R.}~\bibnamefont{Kolodka}},
  \bibinfo{author}{\bibfnamefont{J.}~\bibnamefont{Cibert}}, \bibnamefont{and}
  \bibinfo{author}{\bibfnamefont{H.}~\bibnamefont{Mariette}},
  \bibinfo{journal}{Phys. Rev. Lett.} \textbf{\bibinfo{volume}{102}},
  \bibinfo{pages}{127402} (\bibinfo{year}{2009}).

\bibitem[{\citenamefont{Goryca et~al.}(2009)\citenamefont{Goryca, Kazimierczuk,
  Nawrocki, Golnik, Gaj, Kossacki, Wojnar, and Karczewski}}]{Goryca_PRL_2009}
\bibinfo{author}{\bibfnamefont{M.}~\bibnamefont{Goryca}},
  \bibinfo{author}{\bibfnamefont{T.}~\bibnamefont{Kazimierczuk}},
  \bibinfo{author}{\bibfnamefont{M.}~\bibnamefont{Nawrocki}},
  \bibinfo{author}{\bibfnamefont{A.}~\bibnamefont{Golnik}},
  \bibinfo{author}{\bibfnamefont{J.~A.} \bibnamefont{Gaj}},
  \bibinfo{author}{\bibfnamefont{P.}~\bibnamefont{Kossacki}},
  \bibinfo{author}{\bibfnamefont{P.}~\bibnamefont{Wojnar}}, \bibnamefont{and}
  \bibinfo{author}{\bibfnamefont{G.}~\bibnamefont{Karczewski}},
  \bibinfo{journal}{Phys. Rev. Lett.} \textbf{\bibinfo{volume}{103}},
  \bibinfo{pages}{087401} (\bibinfo{year}{2009}).

\bibitem[{\citenamefont{Reiter et~al.}(2009)\citenamefont{Reiter, Kuhn, and
  Axt}}]{Reiter_PRL_2009}
\bibinfo{author}{\bibfnamefont{D.~E.} \bibnamefont{Reiter}},
  \bibinfo{author}{\bibfnamefont{T.}~\bibnamefont{Kuhn}}, \bibnamefont{and}
  \bibinfo{author}{\bibfnamefont{V.~M.} \bibnamefont{Axt}},
  \bibinfo{journal}{Phys. Rev. Lett.} \textbf{\bibinfo{volume}{102}},
  \bibinfo{pages}{177403} (\bibinfo{year}{2009}).

\bibitem[{\citenamefont{Le~Gall et~al.}(2010)\citenamefont{Le~Gall, Kolodka,
  Cao, Boukari, Mariette, Fern\'andez-Rossier, and Besombes}}]{LeGall_PRB_2010}
\bibinfo{author}{\bibfnamefont{C.}~\bibnamefont{Le~Gall}},
  \bibinfo{author}{\bibfnamefont{R.~S.} \bibnamefont{Kolodka}},
  \bibinfo{author}{\bibfnamefont{C.~L.} \bibnamefont{Cao}},
  \bibinfo{author}{\bibfnamefont{H.}~\bibnamefont{Boukari}},
  \bibinfo{author}{\bibfnamefont{H.}~\bibnamefont{Mariette}},
  \bibinfo{author}{\bibfnamefont{J.}~\bibnamefont{Fern\'andez-Rossier}},
  \bibnamefont{and} \bibinfo{author}{\bibfnamefont{L.}~\bibnamefont{Besombes}},
  \bibinfo{journal}{Phys. Rev. B} \textbf{\bibinfo{volume}{81}},
  \bibinfo{pages}{245315} (\bibinfo{year}{2010}).

\bibitem[{\citenamefont{Baudin et~al.}(2011)\citenamefont{Baudin, Benjamin,
  Lema\^{i}tre, and Krebs}}]{krebs_PRL_2011}
\bibinfo{author}{\bibfnamefont{E.}~\bibnamefont{Baudin}},
  \bibinfo{author}{\bibfnamefont{E.}~\bibnamefont{Benjamin}},
  \bibinfo{author}{\bibfnamefont{A.}~\bibnamefont{Lema\^{i}tre}},
  \bibnamefont{and} \bibinfo{author}{\bibfnamefont{O.}~\bibnamefont{Krebs}},
  \bibinfo{journal}{Phys. Rev. Lett.} \textbf{\bibinfo{volume}{107}},
  \bibinfo{pages}{197402} (\bibinfo{year}{2011}).

\bibitem[{\citenamefont{Oka et~al.}(1999)\citenamefont{Oka, Shen, Takabayashi,
  Takahashi, Mitsu, Souma, and Pittini}}]{Oka_JLum_1999}
\bibinfo{author}{\bibfnamefont{Y.}~\bibnamefont{Oka}},
  \bibinfo{author}{\bibfnamefont{J.}~\bibnamefont{Shen}},
  \bibinfo{author}{\bibfnamefont{K.}~\bibnamefont{Takabayashi}},
  \bibinfo{author}{\bibfnamefont{N.}~\bibnamefont{Takahashi}},
  \bibinfo{author}{\bibfnamefont{H.}~\bibnamefont{Mitsu}},
  \bibinfo{author}{\bibfnamefont{I.}~\bibnamefont{Souma}}, \bibnamefont{and}
  \bibinfo{author}{\bibfnamefont{R.}~\bibnamefont{Pittini}},
  \bibinfo{journal}{J. Lumin.} \textbf{\bibinfo{volume}{83}},
  \bibinfo{pages}{83} (\bibinfo{year}{1999}).

\bibitem[{\citenamefont{Lee et~al.}(2005)\citenamefont{Lee, Dobrowolska, and
  Furdyna}}]{Lee_PRB_2005}
\bibinfo{author}{\bibfnamefont{S.}~\bibnamefont{Lee}},
  \bibinfo{author}{\bibfnamefont{M.}~\bibnamefont{Dobrowolska}},
  \bibnamefont{and} \bibinfo{author}{\bibfnamefont{J.~K.}
  \bibnamefont{Furdyna}}, \bibinfo{journal}{Phys. Rev. B}
  \textbf{\bibinfo{volume}{72}}, \bibinfo{pages}{075320}
  (\bibinfo{year}{2005}).

\bibitem[{\citenamefont{Chekhovich et~al.}(2007)\citenamefont{Chekhovich,
  Brichkin, Chernenko, Kulakovskii, Sedova, Sorokin, and
  Ivanov}}]{Chekhovich_PRB_2007}
\bibinfo{author}{\bibfnamefont{E.~A.} \bibnamefont{Chekhovich}},
  \bibinfo{author}{\bibfnamefont{A.~S.} \bibnamefont{Brichkin}},
  \bibinfo{author}{\bibfnamefont{A.~V.} \bibnamefont{Chernenko}},
  \bibinfo{author}{\bibfnamefont{V.~D.} \bibnamefont{Kulakovskii}},
  \bibinfo{author}{\bibfnamefont{I.~V.} \bibnamefont{Sedova}},
  \bibinfo{author}{\bibfnamefont{S.~V.} \bibnamefont{Sorokin}},
  \bibnamefont{and} \bibinfo{author}{\bibfnamefont{S.~V.}
  \bibnamefont{Ivanov}}, \bibinfo{journal}{Phys. Rev. B}
  \textbf{\bibinfo{volume}{76}}, \bibinfo{pages}{165305}
  (\bibinfo{year}{2007}).

\bibitem[{\citenamefont{Beaulac
  et~al.}(2008{\natexlab{a}})\citenamefont{Beaulac, Archer, van Rijssel,
  Meijerink, and Gamelin}}]{Beaulac_NanoLett_2008}
\bibinfo{author}{\bibfnamefont{R.}~\bibnamefont{Beaulac}},
  \bibinfo{author}{\bibfnamefont{P.~I.} \bibnamefont{Archer}},
  \bibinfo{author}{\bibfnamefont{J.}~\bibnamefont{van Rijssel}},
  \bibinfo{author}{\bibfnamefont{A.}~\bibnamefont{Meijerink}},
  \bibnamefont{and} \bibinfo{author}{\bibfnamefont{D.~R.}
  \bibnamefont{Gamelin}}, \bibinfo{journal}{Nano Lett.}
  \textbf{\bibinfo{volume}{8}}, \bibinfo{pages}{2949}
  (\bibinfo{year}{2008}{\natexlab{a}}).

\bibitem[{\citenamefont{Beaulac
  et~al.}(2008{\natexlab{b}})\citenamefont{Beaulac, Archer, Ochsenbein, and
  Gamelin}}]{Beaulac_AdvFunMat_2008}
\bibinfo{author}{\bibfnamefont{R.}~\bibnamefont{Beaulac}},
  \bibinfo{author}{\bibfnamefont{P.~I.} \bibnamefont{Archer}},
  \bibinfo{author}{\bibfnamefont{S.~T.} \bibnamefont{Ochsenbein}},
  \bibnamefont{and} \bibinfo{author}{\bibfnamefont{D.~R.}
  \bibnamefont{Gamelin}}, \bibinfo{journal}{Advanced Functional Materials}
  \textbf{\bibinfo{volume}{18}}, \bibinfo{pages}{3873}
  (\bibinfo{year}{2008}{\natexlab{b}}).

\bibitem[{\citenamefont{Bussian et~al.}(2009)\citenamefont{Bussian, Crooker,
  Yin, Brynda, Efros, and Klimov}}]{Bussian_NatMat_2009}
\bibinfo{author}{\bibfnamefont{D.~A.} \bibnamefont{Bussian}},
  \bibinfo{author}{\bibfnamefont{S.~A.} \bibnamefont{Crooker}},
  \bibinfo{author}{\bibfnamefont{M.}~\bibnamefont{Yin}},
  \bibinfo{author}{\bibfnamefont{M.}~\bibnamefont{Brynda}},
  \bibinfo{author}{\bibfnamefont{A.~L.} \bibnamefont{Efros}}, \bibnamefont{and}
  \bibinfo{author}{\bibfnamefont{V.~I.} \bibnamefont{Klimov}},
  \bibinfo{journal}{Nature Mater.} \textbf{\bibinfo{volume}{8}},
  \bibinfo{pages}{35} (\bibinfo{year}{2009}).

\bibitem[{\citenamefont{Pacuski et~al.}(2011)\citenamefont{Pacuski,
  Suffczy\'{n}ski, Osewski, Kossacki, Golnik, Gaj, Deparis, Morhain, Chikoidze,
  Dumont et~al.}}]{Pacuski_PRB_2011}
\bibinfo{author}{\bibfnamefont{W.}~\bibnamefont{Pacuski}},
  \bibinfo{author}{\bibfnamefont{J.}~\bibnamefont{Suffczy\'{n}ski}},
  \bibinfo{author}{\bibfnamefont{P.}~\bibnamefont{Osewski}},
  \bibinfo{author}{\bibfnamefont{P.}~\bibnamefont{Kossacki}},
  \bibinfo{author}{\bibfnamefont{A.}~\bibnamefont{Golnik}},
  \bibinfo{author}{\bibfnamefont{J.~A.} \bibnamefont{Gaj}},
  \bibinfo{author}{\bibfnamefont{C.}~\bibnamefont{Deparis}},
  \bibinfo{author}{\bibfnamefont{C.}~\bibnamefont{Morhain}},
  \bibinfo{author}{\bibfnamefont{E.}~\bibnamefont{Chikoidze}},
  \bibinfo{author}{\bibfnamefont{Y.}~\bibnamefont{Dumont}},
  \bibinfo{author}{\bibfnamefont{D.}~\bibnamefont{Ferrand}},
  \bibinfo{author}{\bibfnamefont{J.}~\bibnamefont{Cibert}}, \bibnamefont{and}
  \bibinfo{author}{\bibfnamefont{T.}~\bibnamefont{Dietl}},
  \bibinfo{journal}{Phys. Rev. B} \textbf{\bibinfo{volume}{84}}, 
  \bibinfo{pages}{035214} (\bibinfo{year}{2011}).

\bibitem[{\citenamefont{Papaj et~al.}(2014)\citenamefont{Papaj, Kobak, Rousset,
  Janik, Nawrocki, Kossacki, Golnik, and Pacuski}}]{Papaj_JCG_2014}
\bibinfo{author}{\bibfnamefont{M.}~\bibnamefont{Papaj}},
  \bibinfo{author}{\bibfnamefont{J.}~\bibnamefont{Kobak}},
  \bibinfo{author}{\bibfnamefont{J.}~\bibnamefont{Rousset}},
  \bibinfo{author}{\bibfnamefont{E.}~\bibnamefont{Janik}},
  \bibinfo{author}{\bibfnamefont{M.}~\bibnamefont{Nawrocki}},
  \bibinfo{author}{\bibfnamefont{P.}~\bibnamefont{Kossacki}},
  \bibinfo{author}{\bibfnamefont{A.}~\bibnamefont{Golnik}}, \bibnamefont{and}
  \bibinfo{author}{\bibfnamefont{W.}~\bibnamefont{Pacuski}},
  \bibinfo{journal}{J. Cryst. Growth} \textbf{\bibinfo{volume}{401}},
  \bibinfo{pages}{644 } (\bibinfo{year}{2014}).

\bibitem[{\citenamefont{Krebs et~al.}(2009)\citenamefont{Krebs, Benjamin, and
  Lema\^{i}tre}}]{Krebs_PRB_2009}
\bibinfo{author}{\bibfnamefont{O.}~\bibnamefont{Krebs}},
  \bibinfo{author}{\bibfnamefont{E.}~\bibnamefont{Benjamin}}, \bibnamefont{and}
  \bibinfo{author}{\bibfnamefont{A.}~\bibnamefont{Lema\^{i}tre}},
  \bibinfo{journal}{Phys. Rev. B} \textbf{\bibinfo{volume}{80}},
  \bibinfo{pages}{165315} (\bibinfo{year}{2009}).

\bibitem[{\citenamefont{{Kobak} et~al.}(2014)\citenamefont{{Kobak},
  {Smole{\'n}ski}, {Goryca}, {Papaj}, {Gietka}, {Bogucki}, {Koperski},
  {Rousset}, {Suffczy{\'n}ski}, {Janik} et~al.}}]{Kobak_2014}
\bibinfo{author}{\bibfnamefont{J.}~\bibnamefont{{Kobak}}},
  \bibinfo{author}{\bibfnamefont{T.}~\bibnamefont{{Smole{\'n}ski}}},
  \bibinfo{author}{\bibfnamefont{M.}~\bibnamefont{{Goryca}}},
  \bibinfo{author}{\bibfnamefont{M.}~\bibnamefont{{Papaj}}},
  \bibinfo{author}{\bibfnamefont{K.}~\bibnamefont{{Gietka}}},
  \bibinfo{author}{\bibfnamefont{A.}~\bibnamefont{{Bogucki}}},
  \bibinfo{author}{\bibfnamefont{M.}~\bibnamefont{{Koperski}}},
  \bibinfo{author}{\bibfnamefont{J.-G.} \bibnamefont{{Rousset}}},
  \bibinfo{author}{\bibfnamefont{J.}~\bibnamefont{{Suffczy{\'n}ski}}},
  \bibinfo{author}{\bibfnamefont{E.}~\bibnamefont{{Janik}}},
  \bibinfo{author}{\bibfnamefont{M.}~\bibnamefont{{Nawrocki}}},
  \bibinfo{author}{\bibfnamefont{A.}~\bibnamefont{{Golnik}}},
  \bibinfo{author}{\bibfnamefont{P.}~\bibnamefont{{Kossacki}}}, \bibnamefont{and}
  \bibinfo{author}{\bibfnamefont{W.}~\bibnamefont{{Pacuski}}},
  \bibinfo{journal}{Nat. Comm.} \textbf{\bibinfo{volume}{5}}, 
  \bibinfo{pages}{3191} (\bibinfo{year}{2014}).

\bibitem[{\citenamefont{Goryca et~al.}(2010)\citenamefont{Goryca, Plochocka,
  Kazimierczuk, Wojnar, Karczewski, Gaj, Potemski, and
  Kossacki}}]{Goryca_PRB_2010}
\bibinfo{author}{\bibfnamefont{M.}~\bibnamefont{Goryca}},
  \bibinfo{author}{\bibfnamefont{P.}~\bibnamefont{Plochocka}},
  \bibinfo{author}{\bibfnamefont{T.}~\bibnamefont{Kazimierczuk}},
  \bibinfo{author}{\bibfnamefont{P.}~\bibnamefont{Wojnar}},
  \bibinfo{author}{\bibfnamefont{G.}~\bibnamefont{Karczewski}},
  \bibinfo{author}{\bibfnamefont{J.~A.} \bibnamefont{Gaj}},
  \bibinfo{author}{\bibfnamefont{M.}~\bibnamefont{Potemski}}, \bibnamefont{and}
  \bibinfo{author}{\bibfnamefont{P.}~\bibnamefont{Kossacki}},
  \bibinfo{journal}{Phys. Rev. B} \textbf{\bibinfo{volume}{82}},
  \bibinfo{pages}{165323} (\bibinfo{year}{2010}).

\bibitem[{\citenamefont{Title}(1963)}]{Title_PR_1963}
\bibinfo{author}{\bibfnamefont{R.~S.} \bibnamefont{Title}},
  \bibinfo{journal}{Phys. Rev.} \textbf{\bibinfo{volume}{131}},
  \bibinfo{pages}{2503} (\bibinfo{year}{1963}).

\bibitem[{\citenamefont{Kulakovskii et~al.}(1999)\citenamefont{Kulakovskii,
  Bacher, Weigand, K\"ummell, Forchel, Borovitskaya, Leonardi, and
  Hommel}}]{Hommel_PRL_1999}
\bibinfo{author}{\bibfnamefont{V.~D.} \bibnamefont{Kulakovskii}},
  \bibinfo{author}{\bibfnamefont{G.}~\bibnamefont{Bacher}},
  \bibinfo{author}{\bibfnamefont{R.}~\bibnamefont{Weigand}},
  \bibinfo{author}{\bibfnamefont{T.}~\bibnamefont{K\"ummell}},
  \bibinfo{author}{\bibfnamefont{A.}~\bibnamefont{Forchel}},
  \bibinfo{author}{\bibfnamefont{E.}~\bibnamefont{Borovitskaya}},
  \bibinfo{author}{\bibfnamefont{K.}~\bibnamefont{Leonardi}}, \bibnamefont{and}
  \bibinfo{author}{\bibfnamefont{D.}~\bibnamefont{Hommel}},
  \bibinfo{journal}{Phys. Rev. Lett.} \textbf{\bibinfo{volume}{82}},
  \bibinfo{pages}{1780} (\bibinfo{year}{1999}).

\bibitem[{\citenamefont{Bayer et~al.}(2002)\citenamefont{Bayer, Ortner, Stern,
  Kuther, Gorbunov, Forchel, Hawrylak, Fafard, Hinzer, Reinecke
  et~al.}}]{Bayer_PRB_2002}
\bibinfo{author}{\bibfnamefont{M.}~\bibnamefont{Bayer}},
  \bibinfo{author}{\bibfnamefont{G.}~\bibnamefont{Ortner}},
  \bibinfo{author}{\bibfnamefont{O.}~\bibnamefont{Stern}},
  \bibinfo{author}{\bibfnamefont{A.}~\bibnamefont{Kuther}},
  \bibinfo{author}{\bibfnamefont{A.~A.} \bibnamefont{Gorbunov}},
  \bibinfo{author}{\bibfnamefont{A.}~\bibnamefont{Forchel}},
  \bibinfo{author}{\bibfnamefont{P.}~\bibnamefont{Hawrylak}},
  \bibinfo{author}{\bibfnamefont{S.}~\bibnamefont{Fafard}},
  \bibinfo{author}{\bibfnamefont{K.}~\bibnamefont{Hinzer}},
  \bibinfo{author}{\bibfnamefont{T.~L.} \bibnamefont{Reinecke}},
  \bibinfo{author}{\bibfnamefont{S.~N.} \bibnamefont{Walck}},
  \bibinfo{author}{\bibfnamefont{J.~P.} \bibnamefont{Reithmaier}},
  \bibinfo{author}{\bibfnamefont{F.}~\bibnamefont{Klopf}}, \bibnamefont{and}
  \bibinfo{author}{\bibfnamefont{F.}~\bibnamefont{Sch\"afer}},
  \bibinfo{journal}{Phys. Rev. B} \textbf{\bibinfo{volume}{65}}, 
  \bibinfo{pages}{195315} (\bibinfo{year}{2002}).

\bibitem[{\citenamefont{Akimov et~al.}(2005)\citenamefont{Akimov, Kavokin,
  Hundt, and Henneberger}}]{Akimov_PRB_2005}
\bibinfo{author}{\bibfnamefont{I.~A.} \bibnamefont{Akimov}},
  \bibinfo{author}{\bibfnamefont{K.~V.} \bibnamefont{Kavokin}},
  \bibinfo{author}{\bibfnamefont{A.}~\bibnamefont{Hundt}}, \bibnamefont{and}
  \bibinfo{author}{\bibfnamefont{F.}~\bibnamefont{Henneberger}},
  \bibinfo{journal}{Phys. Rev. B} \textbf{\bibinfo{volume}{71}},
  \bibinfo{pages}{075326} (\bibinfo{year}{2005}).

\bibitem[{\citenamefont{Trojnar et~al.}(2011)\citenamefont{Trojnar,
  Korkusi\ifmmode~\acute{n}\else \'{n}\fi{}ski, Kadantsev, Hawrylak, Goryca,
  Kazimierczuk, Kossacki, Wojnar, and Potemski}}]{Trojnar_PRL_2011}
\bibinfo{author}{\bibfnamefont{A.~H.} \bibnamefont{Trojnar}},
  \bibinfo{author}{\bibfnamefont{M.}~\bibnamefont{Korkusi\ifmmode~\acute{n}\else
  \'{n}\fi{}ski}}, \bibinfo{author}{\bibfnamefont{E.~S.}
  \bibnamefont{Kadantsev}},
  \bibinfo{author}{\bibfnamefont{P.}~\bibnamefont{Hawrylak}},
  \bibinfo{author}{\bibfnamefont{M.}~\bibnamefont{Goryca}},
  \bibinfo{author}{\bibfnamefont{T.}~\bibnamefont{Kazimierczuk}},
  \bibinfo{author}{\bibfnamefont{P.}~\bibnamefont{Kossacki}},
  \bibinfo{author}{\bibfnamefont{P.}~\bibnamefont{Wojnar}}, \bibnamefont{and}
  \bibinfo{author}{\bibfnamefont{M.}~\bibnamefont{Potemski}},
  \bibinfo{journal}{Phys. Rev. Lett.} \textbf{\bibinfo{volume}{107}},
  \bibinfo{pages}{207403} (\bibinfo{year}{2011}).

\bibitem[{\citenamefont{Trojnar et~al.}(2013)\citenamefont{Trojnar,
  Korkusinski, Mendes, Goryca, Koperski, Smolenski, Kossacki, Wojnar, and
  Hawrylak}}]{Trojnar_PRB_2013}
\bibinfo{author}{\bibfnamefont{A.~H.} \bibnamefont{Trojnar}},
  \bibinfo{author}{\bibfnamefont{M.}~\bibnamefont{Korkusinski}},
  \bibinfo{author}{\bibfnamefont{U.~C.} \bibnamefont{Mendes}},
  \bibinfo{author}{\bibfnamefont{M.}~\bibnamefont{Goryca}},
  \bibinfo{author}{\bibfnamefont{M.}~\bibnamefont{Koperski}},
  \bibinfo{author}{\bibfnamefont{T.}~\bibnamefont{Smolenski}},
  \bibinfo{author}{\bibfnamefont{P.}~\bibnamefont{Kossacki}},
  \bibinfo{author}{\bibfnamefont{P.}~\bibnamefont{Wojnar}}, \bibnamefont{and}
  \bibinfo{author}{\bibfnamefont{P.}~\bibnamefont{Hawrylak}},
  \bibinfo{journal}{Phys. Rev. B} \textbf{\bibinfo{volume}{87}},
  \bibinfo{pages}{205311} (\bibinfo{year}{2013}).

\bibitem[{\citenamefont{Besombes and Boukari}(2014)}]{Besombes_PRB_2014}
\bibinfo{author}{\bibfnamefont{L.}~\bibnamefont{Besombes}} \bibnamefont{and}
  \bibinfo{author}{\bibfnamefont{H.}~\bibnamefont{Boukari}},
  \bibinfo{journal}{Phys. Rev. B} \textbf{\bibinfo{volume}{89}},
  \bibinfo{pages}{085315} (\bibinfo{year}{2014}).

\bibitem[{\citenamefont{Arciszewska and
  Nawrocki}(1968)}]{Arciszewska_JChemSol_1966}
\bibinfo{author}{\bibfnamefont{M.}~\bibnamefont{Arciszewska}} \bibnamefont{and}
  \bibinfo{author}{\bibfnamefont{M.}~\bibnamefont{Nawrocki}},
  \bibinfo{journal}{J. Phys. Chem. Solids} \textbf{\bibinfo{volume}{47}},
  \bibinfo{pages}{309} (\bibinfo{year}{1968}).

\bibitem[{\citenamefont{Koudinov et~al.}(2004)\citenamefont{Koudinov, Akimov,
  Kusrayev, and Henneberger}}]{Koudinov_PRB_2004}
\bibinfo{author}{\bibfnamefont{A.~V.} \bibnamefont{Koudinov}},
  \bibinfo{author}{\bibfnamefont{I.~A.} \bibnamefont{Akimov}},
  \bibinfo{author}{\bibfnamefont{Y.~G.} \bibnamefont{Kusrayev}},
  \bibnamefont{and}
  \bibinfo{author}{\bibfnamefont{F.}~\bibnamefont{Henneberger}},
  \bibinfo{journal}{Phys. Rev. B} \textbf{\bibinfo{volume}{70}},
  \bibinfo{pages}{241305} (\bibinfo{year}{2004}).

\bibitem[{\citenamefont{{Kazimierczuk}
  et~al.}(2011)\citenamefont{{Kazimierczuk}, {Smole{\'n}ski}, {Goryca},
  {Klopotowski}, {Wojnar}, {Fronc}, {Golnik}, {Nawrocki}, {Gaj}, and
  {Kossacki}}}]{tkaz_prb_2011}
\bibinfo{author}{\bibfnamefont{T.}~\bibnamefont{{Kazimierczuk}}},
  \bibinfo{author}{\bibfnamefont{T.}~\bibnamefont{{Smole{\'n}ski}}},
  \bibinfo{author}{\bibfnamefont{M.}~\bibnamefont{{Goryca}}},
  \bibinfo{author}{\bibfnamefont{L.}~\bibnamefont{{Klopotowski}}},
  \bibinfo{author}{\bibfnamefont{P.}~\bibnamefont{{Wojnar}}},
  \bibinfo{author}{\bibfnamefont{K.}~\bibnamefont{{Fronc}}},
  \bibinfo{author}{\bibfnamefont{A.}~\bibnamefont{{Golnik}}},
  \bibinfo{author}{\bibfnamefont{M.}~\bibnamefont{{Nawrocki}}},
  \bibinfo{author}{\bibfnamefont{J.~A.} \bibnamefont{{Gaj}}}, \bibnamefont{and}
  \bibinfo{author}{\bibfnamefont{P.}~\bibnamefont{{Kossacki}}},
  \bibinfo{journal}{Phys. Rev. B} \textbf{\bibinfo{volume}{84}},
  \bibinfo{pages}{165319} (\bibinfo{year}{2011}).

\bibitem[{\citenamefont{Smole\ifmmode~\acute{n}\else \'{n}\fi{}ski
  et~al.}(2012)\citenamefont{Smole\ifmmode~\acute{n}\else \'{n}\fi{}ski,
  Kazimierczuk, Goryca, Jakubczyk, K\l{}opotowski, Cywi\ifmmode~\acute{n}\else
  \'{n}\fi{}ski, Wojnar, Golnik, and Kossacki}}]{Smolenski_PRB_2012}
\bibinfo{author}{\bibfnamefont{T.}~\bibnamefont{Smole\ifmmode~\acute{n}\else
  \'{n}\fi{}ski}},
  \bibinfo{author}{\bibfnamefont{T.}~\bibnamefont{Kazimierczuk}},
  \bibinfo{author}{\bibfnamefont{M.}~\bibnamefont{Goryca}},
  \bibinfo{author}{\bibfnamefont{T.}~\bibnamefont{Jakubczyk}},
  \bibinfo{author}{\bibfnamefont{L.}~\bibnamefont{K\l{}opotowski}},
  \bibinfo{author}{\bibfnamefont{L.}~\bibnamefont{Cywi\ifmmode~\acute{n}\else
  \'{n}\fi{}ski}}, \bibinfo{author}{\bibfnamefont{P.}~\bibnamefont{Wojnar}},
  \bibinfo{author}{\bibfnamefont{A.}~\bibnamefont{Golnik}}, \bibnamefont{and}
  \bibinfo{author}{\bibfnamefont{P.}~\bibnamefont{Kossacki}},
  \bibinfo{journal}{Phys. Rev. B} \textbf{\bibinfo{volume}{86}},
  \bibinfo{pages}{241305} (\bibinfo{year}{2012}).

\end{thebibliography}
\end{document}